\documentclass[twocolumn,showpacs,preprintnumbers,amsmath,amssymb]{revtex4}

\usepackage{graphicx}
\usepackage{dcolumn}
\usepackage{bm}
\usepackage{color}
\usepackage{multirow}

\begin{document}
\title{Towards an improved understanding of the relative scintillation efficiency of nuclear recoils in liquid xenon}

\author{A.~Manalaysay}
\email[\emph{Electronic address}: ]{aaronm@physik.uzh.ch}
\affiliation{Physik-Institut, Universit\"{a}t Z\"{u}rich, Z\"{u}rich, 8057, Switzerland}

\date{\today}

\begin{abstract}
Liquid xenon (LXe) particle detectors are a powerful technology in the field of dark matter direct detection, having shown impressive results in recent years and holding strong possibility for leading the field in sensitivity to galactic weakly interacting massive particles (WIMPs) in the future.  The search for WIMPs requires the capability to detect the recoiling nuclei that result when these particles interact with normal matter.  In order to make meaningful statements about an observed signal, or lack thereof, the energy scale of recoiling nuclei in LXe must be known.  Our understanding of this energy scale is contained in a quantity called the relative scintillation efficiency of nuclear recoils, or $\mathcal{L}_{\rm eff}$, and has been studied extensively in the literature, producing seemingly contradictory results.  I examine all the measurements of $\mathcal{L}_{\rm eff}$ that exist, both direct and indirect, and extract the energy dependent behavior that is statistically consistent globally with all values.  Additionally, I examine the measurements covering low energies ($\lesssim$10\,keV, where the largest disagreements exist) and attempt to diagnose the systematic effects that have led to the observed inconsistencies.  I show that virtually all major disparity arises due to efficiency roll-off of the detectors at the low energies, and, when taking this into account, find that the observed behavior of $\mathcal{L}_{\rm eff}$ supports a slowly and smoothly decreasing value with decreasing energy.  Finally, I discuss the prospects for future measurements, and derive a practical limit to what can be achieved.
\end{abstract}


\maketitle
\newcommand{\Leff}{$\mathcal{L}_{\mathrm{eff}}$}
\newcommand{\Kr}{$^{83\mathrm{m}}$Kr~}
\newcommand{\Krns}{$^{83\mathrm{m}}$Kr} 
\newcommand{\Co}{$^{57}$Co~}
\newcommand{\Cons}{$^{57}$Co}
\newcommand{\Rb}{$^{83}$Rb~}
\newcommand{\Rbns}{$^{83}$Rb}

\section{Introduction}
\label{sec:intro}
A significant amount of attention has been placed in recent years on the use of liquid xenon (LXe) as a medium for particle detection.  In particular, LXe particle detectors have stirred much interest in the field of dark matter direct detection \cite{Angle:2007uj,Lebedenko:2008gb,Aprile:2010um}, experiments that search for Weakly Interacting Massive Particles (WIMPs) leftover from the early universe.  In such studies, the interaction of WIMPs in the detector would result in recoiling Xe nuclei of typically less than 100\,keV \cite{Bertone:2004pz}.

As a particle loses energy in LXe, it leaves behind a track of ionized and electrically excited Xe atoms (collectively referred to as electronic excitation).  Some free electrons liberated by this process can be drifted away from the interaction site and collected for measurement.  Additionally, the remaining electrons may recombine with Xe ions, and together with the excited Xe atoms will de-excite in a process that produces 178\,nm scintillation photons \cite{Jortner:1972}.  These photons can be detected by photomultiplier tubes (PMTs) and converted to an equivalent energy by a quantity called the light yield, which gives the number of photoelectrons (pe) expected per unit of deposited energy.

It is difficult to predict the light yield in absolute terms as it depends on a number of factors, including the energy and identity of the particle.  \Leff, or the relative scintillation efficiency of nuclear recoils, is defined as the light yield of nuclear recoils relative to that of 122\,keV gamma rays from \Cons.  The light yield depends on the electronic stopping power ($dE/dx$), which depends on the energy of the particle, and therefore \Leff~is an energy dependent quantity.  The deviation of \Leff~from unity has several causes.  First, while recoiling electrons lose virtually all of their energy through electronic excitation, a significant amount of energy lost by recoiling nuclei goes into heat, and can therefore never be collected as ionization or scintillation \cite{lindhard:63}.  Second, some of the energy going into electronic excitation can be lost by bi-excitonic quenching which removes a fraction of the scintillation photons.  The amount of energy lost in this manner depends on the exciton density in the recoil track, which differs greatly for electronic and nuclear recoils \cite{Hitachi:2005ti}.  Third, a fraction of the ionized electrons are carried away from the recoil track by thermal motion, preventing them from recombining with ions to produce scintillation photons, and results in a reduced scintillation signal \cite{Doke2002}.  The fraction of lost electrons depends on the ionization density of the recoil track, again differing for electronic and nuclear recoils.  Fourth, the scintillation yield of electronic recoils is also energy dependent, and therefore the energy scale obtained from a calibration at 122\,keV is not necessarily valid for electronic recoils lower energies \cite{Manalaysay:2009yq}.  Fifth, the ratio of excited xenon atoms to ionized xenon atoms ($N_{\rm ex}/N_{\rm ion}$) differs greatly for electronic and nuclear recoils \cite{CEDahl}.

Measurements of \Leff~are performed in two manners, one direct, the other indirect.  The direct method uses neutron spectroscopy; a beam of monoenergetic neutrons is incident upon a LXe detector where it scatters, and some of the neutrons scattering under an angle, $\theta$, are tagged by an organic scintillator capable of neutron-gamma discrimination based on pulse shape discrimination (PSD).  The true energy of the recoil is given by the kinematics alone,
\begin{equation} \label{eq:N_scatter_E}
E_r \approx 2E_n \frac{m_nM_{N}}{(m_n+M_{N})^2}(1 - \mathrm{cos}\,\theta),
\end{equation}
where $E_r$ is the energy of the recoiling Xe nucleus, $E_n$ is the energy of the incoming neutron, and $m_n$ and $M_N$ are the masses of the neutron and nucleus, respectively.  The approximation is valid when $M_N\gg m_n$ and $E_n\ll m_nc^2$.  When the tagging scintillator is placed at a certain angle, one therefore knows the recoil energy and measures the LXe response to those events.  The indirect method uses a continuum-energy source of neutrons and compares the resulting spectrum observed in the LXe with a predicted spectrum from a Monte Carlo (MC) simulation.  The behavior of \Leff~is then determined by an iterative best-fitting procedure, each time scaling the MC spectrum with a different \Leff~model until the best fit is achieved with the actual data.

Though many direct measurements of \Leff~exist in the literature \cite{Arneodo:2000vc,Bernabei:01,Akimov:02,Aprile:05,Chepel:06,Aprile:2008rc,Manzur:2009hp}, the LXe WIMP search experiments have become increasingly focused upon low-energy interactions, in regions of energy where only three of these measurements provide coverage (below ~10\,keV) \cite{Chepel:06,Aprile:2008rc,Manzur:2009hp}.  This push to lower and lower energies is due to the fact that WIMPs are expected to produce an exponentially decreasing differential spectrum as a function of energy, and therefore a small improvement in energy threshold yields a large improvement in overall WIMP sensitivity.  This improvement in sensitivity comes at a cost, however, because the nuclear recoil energy threshold of a LXe detector can be known only insofar as \Leff~is known.  Were the three measurements of \Leff~below 10\,keV consistent and precise, no discussion of this sort would be necessary.  They are, instead, not consistent, and it is this uncertainty of \Leff~at low energies that provides the largest systematic uncertainty to the results reported in the LXe WIMP searches.

\begin{figure}[h!]
		\includegraphics[width=.48\textwidth]{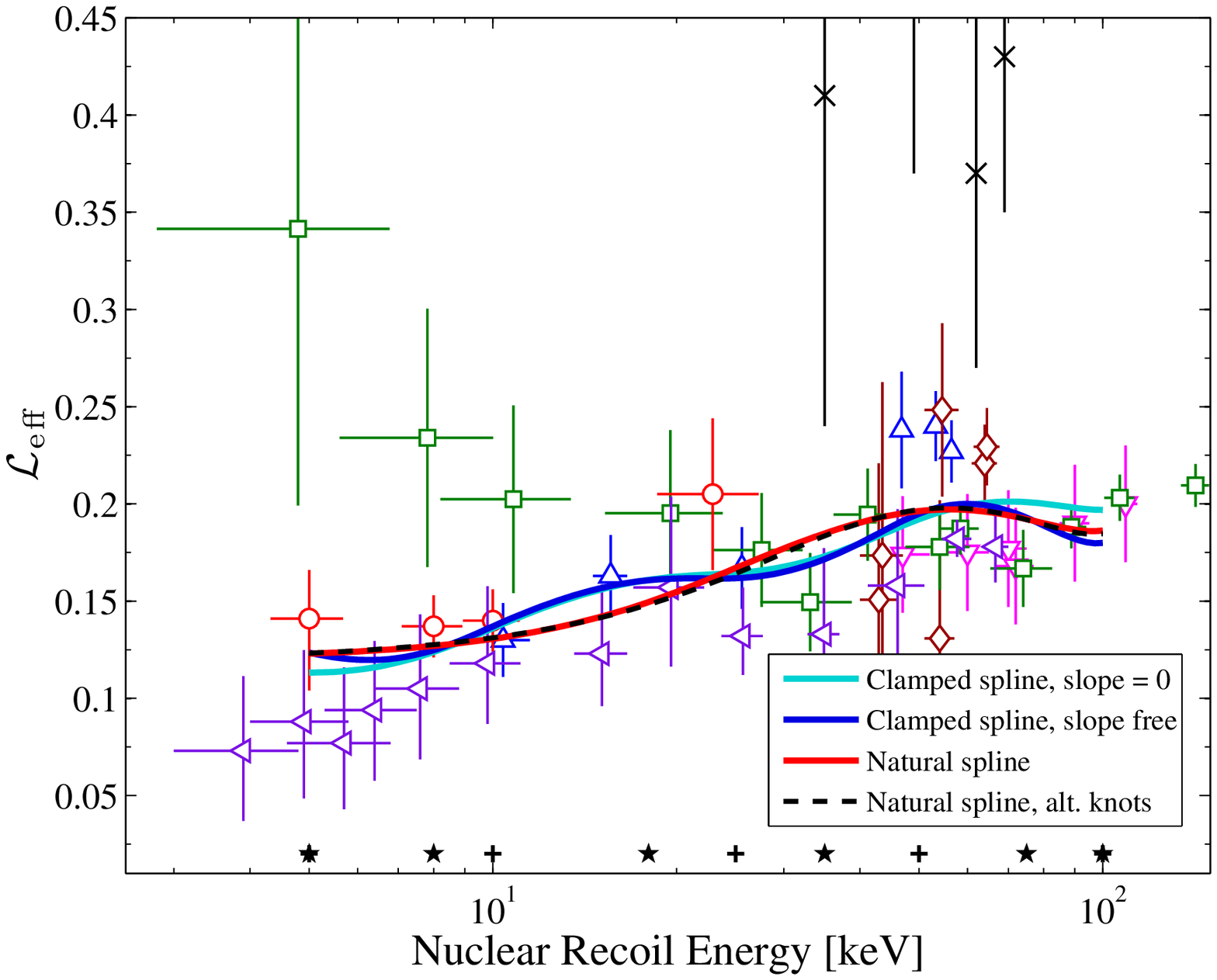} 
	\caption{Measured \Leff~values as a function of Xe nuclear recoil energy.  Symbols 
	correspond to 
	(\textcolor[rgb]{.4784,.0627,.8941}{$\lhd$})--Manzur \emph{et al.} \cite{Manzur:2009hp};
	(\textcolor{red}{$\circ$})--Aprile \emph{et al.} (2009) \cite{Aprile:2008rc};
	(\textcolor[rgb]{0,.5,0}{\bf$\square$})--Chepel \emph{et al.} \cite{Chepel:06};
	(\textcolor{blue}{$\bigtriangleup$})--Aprile \emph{et al.} (2005) \cite{Aprile:05};
	(\textcolor[rgb]{.6,0,0}{$\lozenge$})--Akimov \emph{et al.} \cite{Akimov:02};
	($\times$)--Bernabei et al.~\cite{Bernabei:01};
	(\textcolor[rgb]{1.,0,1.}{$\bigtriangledown$})--Arneodo \emph{et al.} \cite{Arneodo:2000vc}.  A best fit is performed according to four different methods (see text), three using the spline knots placed at energies indicated by a black `+', and one with alternate spline knots indicated by a black `$\star$'.}
	\label{fig:Leff_fits}
\end{figure}

I attack the problem here in two ways.  Section \ref{sec:best_fit} presents a ``fair'' approach, considering every direct measurement to be equal in validity.  Results are weighted only with their reported error bars and the global \Leff~behavior that is statistically consistent with all measurements is extracted.  Section \ref{sec:beam_comparison} critiques the three direct measurements in the energy region most relevant to WIMP searches, while section \ref{sec:indirect_comparison} examines the two indirect measurements.  I must point out that I am a co-author on two of these results, one direct \cite{Aprile:2008rc}, one indirect \cite{Sorensen:2008ec}, so I run the risk of giving a biased critique.  However, fully aware of this risk, I claim no intentional bias, and have made every effort to prevent possible unintentional proclivity.

\section{Global Behavior of \Leff}
\label{sec:best_fit}
In an effort to reconcile the various \Leff~measurements, a global fit is performed here to all reported direct measurements.  Such a treatment is the most unbiased approach possible, with each measured value weighted by its reported total uncertainty.  \Leff~is parameterized as a function of recoil energy by a cubic spline interpolation with spline knots placed at fixed recoil energies.  The spline interpolates between knots using cubic polynomials, requiring continuity of the function, and its first and second derivatives.  The free parameters of the fit are taken as the value of \Leff~at each spline knot.  The energies of the spline knots, 5, 10, 25, 50, and 100\,keV, are chosen to be at locations where there are at least three literature measurements in close proximity, with the exception of the knot at 100\,keV as only two measurements exist in this range.

\begin{figure*}[htp!]
		\includegraphics[width=.48\textwidth]{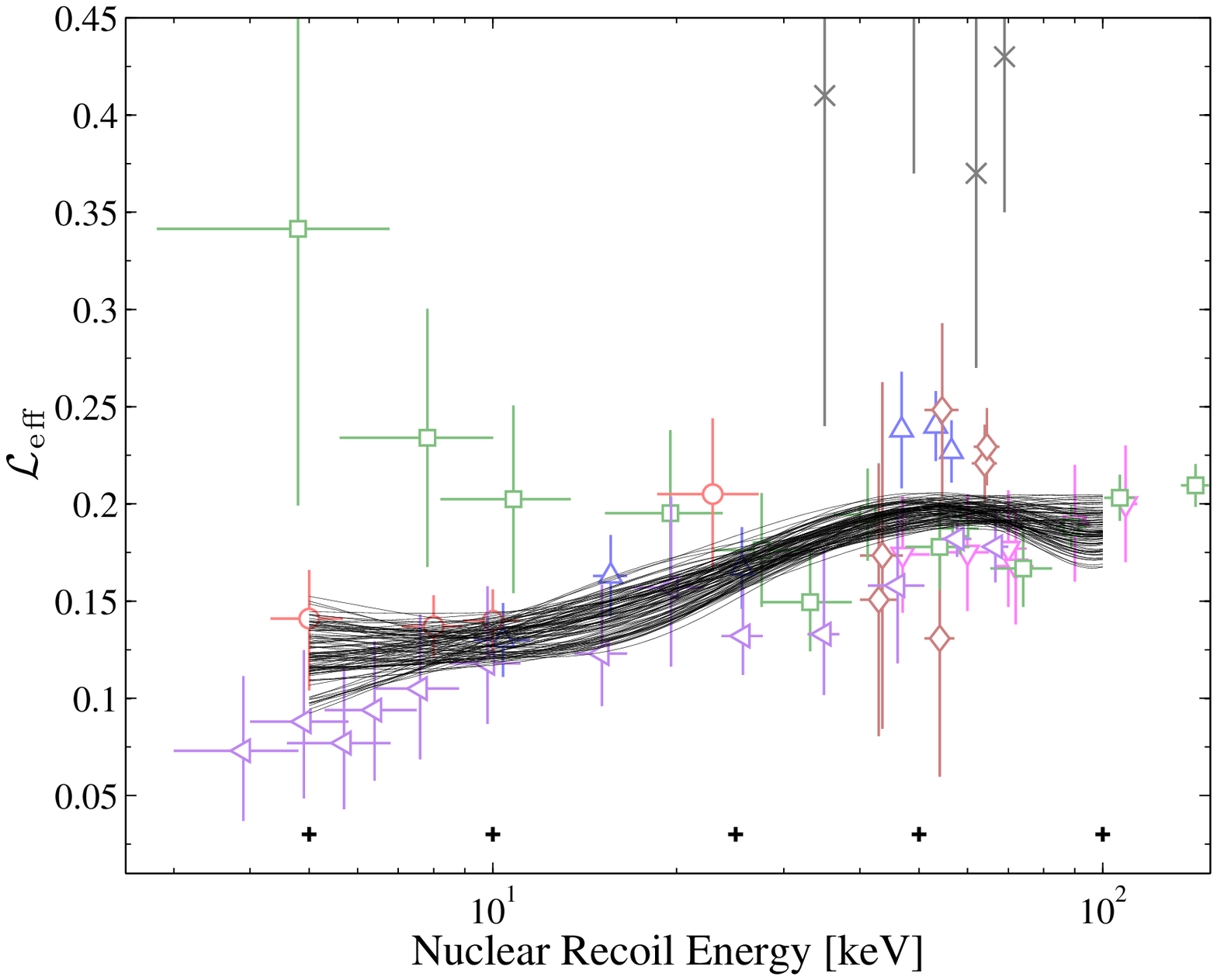} ~~~
		\includegraphics[width=.48\textwidth]{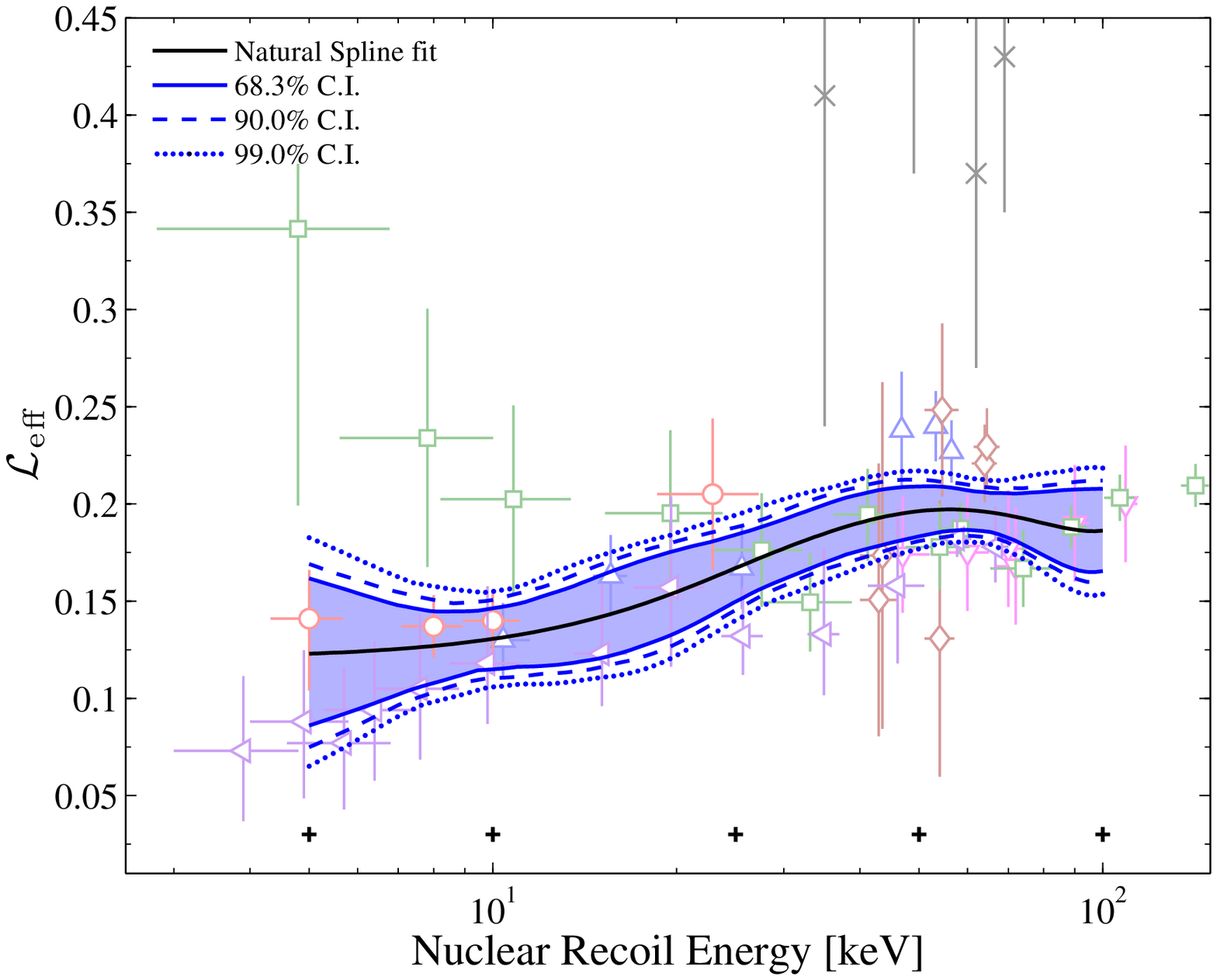}
	\caption{(\emph{Left}) An example of 100 curves allowed at the 68.3\% confidence level.  (\emph{Right}) The 68.3\%, 90.0\%, and 99.0\% confidence intervals allowed by all beam measurements in the literature.}
	\label{fig:Leff_CI}
\end{figure*}

The cubic spline is chosen as a parameterization because it permits a large degree of variability in \Leff, and does not assume \emph{a priori} any particular functional shape.  In order to calculate a cubic spline, boundary conditions must be chosen which govern the behavior of the interpolation at the endpoints.  A \emph{natural cubic spline} forces the interpolation function, $f$, to have null second derivatives at the endpoints: $f''(x_{\rm first})=f''(x_{\rm last})=0$, where $x_{\rm first}$ is the first spline knot, and $x_{\rm last}$ is the last spline knot.  A \emph{clamped cubic spline} fixes, instead, the first derivatives at the endpoints: $f'(x_{\rm first}) = a, f'(x_{\rm last})=b$, where $a$ and $b$ are arbitrary values.
In order to account for these possibilities, the fit is performed under three different conditions, seen in Figure \ref{fig:Leff_fits}: (1) a clamped spline with the slope at the endpoints fixed at zero ($\chi^2/\rm{ndf}=1.59$); (2) a clamped spline with the slope at the endpoints left as two additional free fit parameters ($\chi^2/\rm{ndf}=1.55$); (3) a natural spline ($\chi^2/\rm{ndf}=1.57$).  The values of \Leff~at the spline knots for these three conditions are shown in Table \ref{tab:spline_points}.  A fourth fit ($\chi^2/\rm{ndf}=1.53$) is performed with a natural spline using an alternate set of knot energies (5, 8, 18, 35, 75, and 100\,keV).  In all cases, the behavior of \Leff~outside of the knot endpoints is extrapolated as a line whose slope is equal to the slope at that endpoint. 

\begin{table}[htp!]
\begin{center}
\caption{\label{tab:spline_points}The best fit values of the spline fits displayed in Figure \ref{fig:Leff_fits}.  Results are shown for a natural spline (``nat.''), clamped spline with slope fixed at 0 (``clamp,0''), and clamped spline with endpoint slopes left as free parameters (``clamp,f'').}
	\begin{tabular*}{0.48\textwidth}{@{\extracolsep{\fill}} c | c | c | c }
		\hline \hline 
		$E$\,(keV) & \Leff~(nat.) & \Leff~(clamp,0) & \Leff~(clamp,f)\\
		\hline
		5   & 0.123  &  0.113 & 0.124 \\ 
		10  & 0.131  &  0.136 & 0.137 \\ 
		25  & 0.167  &  0.164 & 0.162 \\
		50  & 0.196  &  0.192 & 0.195 \\
		100 & 0.186  &  0.197 & 0.180 \\
		\hline \hline
	\end{tabular*}
\end{center}
\end{table}

The four curves in Figure \ref{fig:Leff_fits} are calculated in order to estimate the uncertainty introduced by different choices of interpolation.  For a single spline choice (natural spline), the uncertainty is determined by the MC method.  The procedure choses one random \Leff~value for each spline knot, interpolates, and the resulting $\chi^2$ value is calculated.  These steps are carried out $5\times10^6$ times, each time choosing a new set of random numbers.  Confidence intervals (C.I.) are determined by those models giving $\chi^2<\chi^2_{\rm min}+Q_{\gamma}$, where $Q_{\gamma}$ = 5.89, 9.24, and 15.1 for 68.3\%, 90.0\%, and 99.0\% C.I., respectively, for five free parameters \cite{Cowan:98}.  An example of 100 models within the 68.3\% confidence interval are shown in Figure \ref{fig:Leff_CI} (\emph{left}).  The full area covered by all allowed models in the MC is shown in Figure \ref{fig:Leff_CI} (\emph{right}) for the three confidence levels mentioned here.  At 5\,keV, the 68.3\%, 90.0\%, and 99.0\% C.I.s allow for variation in \Leff~of $\pm$30\%, $\pm$38\%, and $\pm$48\%, respectively.  The same C.I.s allow, at 10\,keV, variation in \Leff~of $\pm$12\%, $\pm$15\%, and $\pm$19\%, respectively.

\section{Comparison of Direct Measurements}
\label{sec:beam_comparison}
\begin{figure}[htp]
		\includegraphics[width=.45\textwidth]{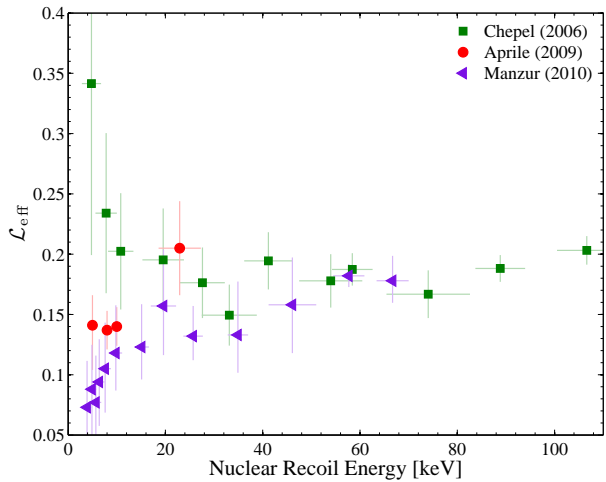}
	\caption{Results of the three direct \Leff~measurements that extend below 10\,keV, shown alone to highlight the disagreement at low energies.}
	\label{fig:threeBlw10}
\end{figure}

The analysis performed in section \ref{sec:best_fit} represents the true global trend of all reported beam measurements, and suggests a slowly decreasing value of \Leff~below $\sim$50\,keV.  However, though this method is fair in that it weights measurements only by their reported error bars, it has no power to make conclusions regarding the true behavior of \Leff~in the lowest energies.  Theoretical attempts have been made to understand the energy dependent behavior of \Leff~\cite{Hitachi:2005ti,Manzur:2009hp}.  However, these studies are based on Lindhard theory \cite{lindhard:63} of nuclear quenching which is invalid in LXe below 10\,keV \cite{Hitachi:2007zz,Mangiarotti:2006ye}, and we must instead rely on measurements alone.  The discrepancy between the three measurements below 10\,keV has yet no explanation.  Seen in Figure \ref{fig:threeBlw10}, rising \Leff~behavior (for decreasing energies) below $\sim$30\,keV is supported by Chepel \cite{Chepel:06}; the results from Aprile \cite{Aprile:2008rc} are consistent with a flat, or slowly decreasing \Leff~below 10\,keV; Manzur \cite{Manzur:2009hp} shows a slowly falling \Leff~until $\sim$10\,keV, followed by a sharp drop below this energy.  To address this issue, a thorough comparison of the various strengths and weaknesses of each measurement is here made.  Only after such a comparison can one make any reasonable statements about which of the three behaviors most closely represents reality.

\begin{table*}[htp!]
\begin{center}
\begin{minipage}[b]{1.0\linewidth}
\caption{\label{tab:sys_comparison}Comparison of the systemic properties of Chepel\,\cite{Chepel:06}, Aprile\,\cite{Aprile:2008rc}, and Manzur\,\cite{Manzur:2009hp}, the three measurements of \Leff~below 10\,keV.  Row labels indicate: (`n-gen') process for neutron generation (projectile $\rightarrow$ target), D, T, and P stand for deuterium, tritium, and protons, respectively.; (`$E_n$') reported energy of neutrons and energy spread; (`Collim.?') whether or not the neutron beam was collimated; (`Photo cov.') fraction of LXe volume surface covered by photodetectors; (`LXe dia.') diameter of LXe volume; (`nScint.~dia.') diameter of organic tagging scintillator; (`LXe-nScint') distance between LXe volume and tagging scintillator; (`$\Delta_{\theta}$ @45$^{\circ}$') range of possible scattering angles for a measurement at 45$^{\circ}$; (`$L_y$') light yield from $^{57}$Co; (`$^{57}$Co $E$-res.') energy resolution of the $^{57}$Co peak, as $\sigma$/$\mu$; ('Res.~acc.?) whether or not the analysis considered the uncertainty in the energy resolution; (`$\epsilon$\,(5\,keV)') the detection efficiency for the measurement of 5\,keV nuclear recoils; (`Effic.~acc.?') whether or not the analysis accounted for the detection efficiency in LXe.}
	\begin{tabular*}{1.0\textwidth}{@{\extracolsep{\fill}}l | c | c | c | c | c | c | c | c | c | c | c | c | c }
		\hline \hline
		\multirow{2}{*}{Study} &
		\multirow{2}{*}{n-gen} & 
		$E_n$ & 
		\multirow{2}{*}{Collim.?} &
		Photo & 
		LXe & 
		nScint &
		LXe- &
		$\Delta_{\theta}$ &
		$L_y$ & 
		$^{57}$Co &
		Res. &
		\multirow{2}{*}{$\epsilon$\,(5\,keV)}	 &
		Effic. \\
		 & & (MeV) & & cov.& dia.(cm)&dia.(cm)&nScint&@45$^{\circ}$&(pe/keV)&$E$-res.& acc.? & &acc.?\\
		\hline
		\multirow{2}{*}{Chepel (2006)}&
		\multirow{2}{*}{D$\rightarrow$D}&
		\multirow{2}{*}{6-8$\pm$1\%}&
		\multirow{2}{*}{yes}&
		\multirow{2}{*}{20\%}&
		\multirow{2}{*}{16.5}&
		\multirow{2}{*}{n/a}&
		\multirow{2}{*}{1\,m}&
		\multirow{2}{*}{10$^{\circ}$\footnote{\,Assuming nScint dia. is 7.6\,cm.}}&
		\multirow{2}{*}{5.5}&
		\multirow{2}{*}{7.6\%}&
		\multirow{2}{*}{no} &
		\multirow{2}{*}{$\sim$30\%\footnote{\,Estimated based on $L_y$.}} &
		\multirow{2}{*}{no} \\
		 & & & & & & & & & & & & & \\
		\hline
		\multirow{2}{*}{Aprile (2009)}&
		\multirow{2}{*}{P$\rightarrow$T}&
		\multirow{2}{*}{1.0$\pm$8\%}&
		\multirow{2}{*}{no}&
		\multirow{2}{*}{95\%}&
		\multirow{2}{*}{2.5}&
		\multirow{2}{*}{7.6}&
		\multirow{2}{*}{50\,cm}&
		\multirow{2}{*}{9$^{\circ}$}&
		\multirow{2}{*}{20}&
		\multirow{2}{*}{9.0\%}&
		\multirow{2}{*}{no} &
		\multirow{2}{*}{97\%}&
		\multirow{2}{*}{yes} \\
		 & & & & & & & & & & & & & \\
		\hline
		\multirow{2}{*}{Manzur (2010)}&
		\multirow{2}{*}{D$\rightarrow$D}&
		\multirow{2}{*}{2.8\footnote{\,The energy spread, not reported in \cite{Manzur:2009hp}, is estimated by the manufacturer \cite{Chichester:2006,SimpsonJ_DDgen_private} of the neutron generator used in this study to be $\pm$8\%.  Additionally, due to the thickness of the target and the angular acceptance, the mean neutron energy is $\sim$2.6\,MeV (see text).}}&
		\multirow{2}{*}{no}&
		\multirow{2}{*}{56\%}&
		\multirow{2}{*}{5.0}&
		\multirow{2}{*}{7.6\footnote{\,Incorrectly reported in \cite{Manzur:2009hp} as 3.8\,cm \cite{Manzur_private}.  }} &
		12\,cm\footnote{\,The distance between LXe and nScint.~varied in this range depending on the available space \cite{Manzur_private}.}&
		46$^{\circ}$&
		\multirow{2}{*}{11}&
		\multirow{2}{*}{8.8\%}&
		\multirow{2}{*}{yes} &
		\multirow{2}{*}{44\%}&
		\multirow{2}{*}{yes} \\
		 & & & & & & &20\,cm &26$^{\circ}$ & & & & & \\		
		\hline \hline
	\end{tabular*}
\end{minipage}
\end{center}
\end{table*}
At first glance, measurement of \Leff~in a beam study is simple.  By tagging neutrons that scatter only under a chosen angle, a peaked scintillation spectrum in the LXe detector is produced.  A calibration of the detector with $^{57}$Co (122\,keV gamma rays) determines the detector's light yield, $L_y$, which measures how many photoelectrons (pe) are collected from the photomultiplier tubes (PMTs) for a given energy deposition.  The scintillation signal, in pe, from the tagged neutron data can be converted to an energy in `keVee', or `keV, electronic recoil equivalent', by scaling the signal with $L_y$.  \Leff~is then simply found by the ratio of the observed electronic recoil equivalent energy to the true recoil energy:
\begin{equation} \label{eq:LeffDef}
\mathcal{L}_{\rm eff} = \frac{E_{ee}}{E_{nr}} ,
\end{equation}
where $E_{ee}$ is the observed electronic recoil equivalent energy (based on 122\,keV gamma rays), and $E_{nr}$ is the true energy of the nuclear recoil (fixed by the kinematics).  This relation holds only for zero applied electric fields, but inclusion of non-zero fields can be done by multiplying the right hand side of equation \ref{eq:LeffDef} by $S_e/S_n$,  where $S_{e(n)}$ is the field quenching factor for electronic (nuclear) recoils.  Systematic uncertainties, therefore, fall into two categories: those affecting $E_{nr}$, and those affecting $E_{ee}$.  Table \ref{tab:sys_comparison} compares many of the properties of each experiment that can lead to significant systematic uncertainties.

\subsection{Systematics affecting $E_{nr}$}
\label{sec:beam_comparison:Enr}

The origin of systematic uncertainties in $E_{nr}$ can be found by examining equation \ref{eq:N_scatter_E}.  For neutrons generated by a beam, the resulting neutron energy, $E_n$, can have a spread due to the thickness of the target.  This is because the projectile (deuterons in \cite{Chepel:06} and \cite{Manzur:2009hp}, protons in \cite{Aprile:2008rc}) can lose energy in the target before producing a neutron.  The neutron energy, and energy spread, are listed in Table \ref{tab:sys_comparison}.  The fraction of energy lost by the projectile in the target decreases with increasing projectile energy, which explains why the neutron energy of Chepel is more tightly constrained than that of Aprile and Manzur.  The energy of the neutrons reported in \cite{Manzur:2009hp} (2.8\,MeV) is valid for forward neutrons produced by 80\,keV deuterons incident upon deuterium; however, the thickness of the target is much larger than the penetration depth of the deuterons \cite{SimpsonJ_DDgen_private}.  This means that 2.8\,MeV represents the \emph{maximum} neutron energy, with the average energy lying below this value.  Additionally, neutrons produced at other angles (lower energies) can scatter to the forward direction, thus bringing the average neutron energy down even further, to around 2.6\,MeV.  A lower true neutron energy in equation \ref{eq:N_scatter_E} translates into a lower true recoil energy, which in turn translates through equation \ref{eq:LeffDef} to systematically higher \Leff~values.  This could partially explain why the Manzur data points appear to be systematically lower (at all energies) than other measurements.

A spread in $E_{nr}$ can also arise due to the finite size of the LXe and tagging detectors.  Though the tagging scintillator is placed at a fixed scattering angle, this angle is the \emph{true} scattering angle only for those neutrons scattering in the center of the LXe, and in the tagging scintillator along the axis connecting its center to the LXe's center.  In reality, neutrons may scatter anywhere in the volume of both detectors, yielding a range of true scattering angles, and thus, a range of true recoil energies.  The degree to which this range is confined depends on the size of both detectors, and on the distance of separation between the two detectors.  These parameters are compared in the eighth column of Table \ref{tab:sys_comparison}.  Also listed is the range of scattering angles allowed when the tagging scintillator is placed at 45$^{\circ}$.  As the size of the tagging scintillator used by Chepel is not mentioned in their paper, I have here assumed it to be 3 inches (7.6\,cm) in diameter, typical of this sort of detector.

Additional systematic effects pertaining to the true recoil energy involve non-peaked backgrounds coming from events with multiple interaction vertices.  Such an event can be one in which the neutron scatters multiple times in the LXe (``multiple scatter''), or one in which the neutron scatters once in the LXe and once in any of the detector materials surrounding the active LXe volume (``materials scatter'').  Multiple scatters are minimized by reducing the LXe volume (as done by Aprile), while materials scatters are controlled by minimizing the mass of detector materials (as done by Manzur).  The detector used by Aprile contained large amounts of PTFE surrounding the active LXe volume.  This was used in order to fill the space between the active LXe and the walls of the much larger chamber that would otherwise be filled with passive LXe (``passive'', meaning it is not viewed by PMTs).  Both materials, PTFE and passive LXe, can present additional scattering points for neutrons before and after they scatter in the active LXe.  However, given the lower mass of nuclei in PTFE (F--19\,u, C--12\,u) than Xe (131\,u), neutrons tend to lose much more energy through scatters in PTFE than in LXe.  This lowered energy can then usually be identified as an increased time of flight (ToF) between LXe and tagging scintillator.  Nevertheless, the large amounts of PTFE used introduced considerable backgrounds in the results of \cite{Aprile:2008rc}, even after cutting events with large ToF.  These backgrounds were estimated by MC simulations and subtracted, but the uncertainty in this background estimate contributed significantly to the total uncertainties reported.  

\subsection{Systematics affecting $E_{ee}$}
\label{sec:beam_comparison:Eee}

Systematic processes affecting the observed energy, $E_{ee}$, are due to the properties of the LXe detector itself.  The energy resolution of the detector can cause a true energy peak to appear skewed, because signals on the low energy side of the peak experience proportionally larger Poisson fluctuations than those on the high energy side.  This effect generally pulls the measured \Leff~lower than the true value (since low-energy recoils experience proportionally larger fluctuations than the high-energy recoils), and is highly dependent on the observed number of pe.  For example, the light yields, $L_y$, of the detectors used by Chepel, Aprile, and Manzur are 5.5, 20, and 11 pe/keVee, respectively.  For nuclear recoils of 5\,keV, the energy resolution (assuming Poisson fluctuations on the number of pe and amplification fluctuations of the PMTs) gives a systematic error on the calculated \Leff~in these studies of roughly 40\%, 0.1\%, and 5\%, respectively, in the case that the true value of \Leff~at this energy is 0.123 (here, I have assumed that the spread in $E_{nr}$ is given by the horizontal error bars of the 5\,keV point from each measurement in Figure \ref{fig:threeBlw10}).  Deviations of the pe fluctuations from Poissonianity can change these errors by factors of a few.  Additionally, the effect is worsened if the spread in the true recoil energy, $E_{nr}$, is large, and also if the true value of \Leff~is changing rapidly within that spread.

\begin{figure}[htp!]
		\includegraphics[width=.46\textwidth]{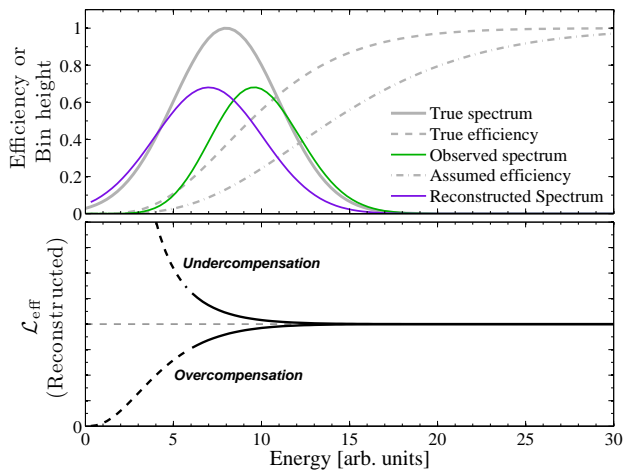} 
	\caption{(\emph{Top}) An example of how detection efficiency affects the position of a peak.  The solid grey curve is a hypothetical true peak, but due to a falling trigger efficiency in this region (grey dashed) it is observed as the green curve, whose peak is shifted to higher energies.  If the experimenter attempts to compensate the spectrum with an incorrect trigger efficiency (grey dot-dashed), the resulting peak (purple) will be shifted below the true value.  (\emph{Bottom}) For the hypothetical situation where the true \Leff~behavior is flat with energy (grey dashed), undercompensation and overcompensation for the total detection efficiency results in a reconstructed \Leff~that rises and falls, respectively, with decreasing energy.}
	\label{fig:Leff_Eff_Reconstruct_ex}
\end{figure}

\subsubsection*{Detection Efficiency}
\label{sec:beam_comparison:Eee:efficiency}

A more significant effect results from the total photon detection efficiency of the detector.  This total detection efficiency is the combination of the geometrical light collection efficiency (LCE), quantum efficiency (QE) of the PMT photocathodes, collection efficiency of the first dynode, PMT amplification effects, hardware trigger efficiency, and software acceptance.  The geometrical LCE, QE, and the hardware trigger efficiency usually provide the dominant contributions.  The first efficiency (LCE) varies depending on where in the LXe volume the energy is deposited.  Because the spatial distribution from $^{57}$Co gamma rays differs from that of neutrons, a systematic shift in the reconstructed \Leff~value can arise in situations where the spatial distribution of the two classes of events differs greatly.  For such small detectors as used in these three studies, this effect from spatially-varying light yields is minimal.

At low energies, the total detection efficiency produces a much more significant effect.  Due to, for example, Poisson fluctuations on the number of detected photoelectrons, a low energy event will have a probability of being detected that is less than unity.  Figure \ref{fig:Leff_Eff_Reconstruct_ex} (\emph{top}) shows how such a detection efficiency ``roll-off'' can affect the position of the observed spectrum.  If this detection efficiency is strongly changing in the region of the peak, the observed peak will be shifted to higher energies than the true peak (green curve).  As the true recoil energy decreases, the observed peak therefore stays at a relatively constant value.  It is clear, then, from equation \ref{eq:LeffDef} that such an effect (decreasing $E_{nr}$ with constant $E_{ee}$) results in a reconstructed \Leff~value that rises at low energies.  Figure \ref{fig:Leff_Eff_Reconstruct_ex} (\emph{bottom}) shows the resulting \Leff~in the case that this detection efficiency is either unaccounted---or undercompensated for.  

\begin{figure}[htp!]
		\includegraphics[width=.46\textwidth]{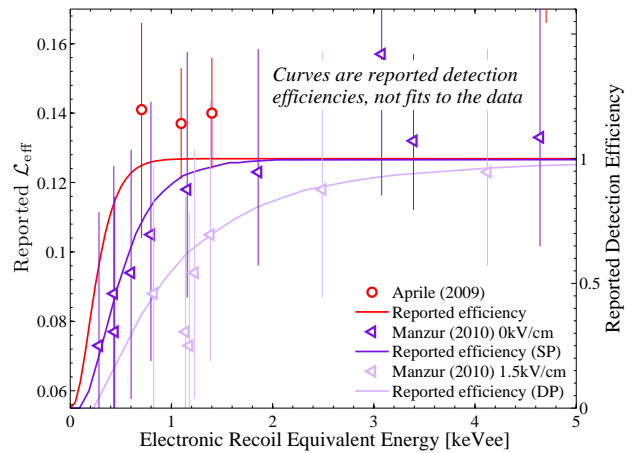}
	\caption{The reported \Leff~values from Aprile\,\cite{Aprile:2008rc} (red circles) and Manzur\,\cite{Manzur:2009hp} zero field (dark purple triangles) and 1.5\,kV/cm (light purple triangles) as a function of electronic recoil equivalent energy.  Superimposed are the detection efficiencies of each study.  The trigger and acquisition used by Manzur differed for single phase (SP) and dual phase (DP).}
	\label{fig:Leff_Eff_Reconstruct}
\end{figure}

Because the roll-off of the detection efficiency is related to the probability to detect a photon, the energy scale at which this roll-off becomes important is related to the light yield, $L_y$.  One therefore expects that the detector used by Chepel, having the lowest value of $L_y$, to show the strongest effects due to the efficiency roll-off.  Though Chepel's detection efficiency is not mentioned in \cite{Chepel:06} or \cite{Neves:2005}, it can be roughly estimated by scaling Manzur's efficiency by $L_y$.  This estimation predicts that the efficiency roll-off becomes significant at around 2\,keVee, or for measurements with recoil energies less than $\sim$12\,keV.  It is precisely at this energy where the \Leff~values of Chepel begin to rise significantly.  One can therefore conclude that the rise in \Leff~observed by Chepel \emph{et al}.~is non-physical, and is instead an artifact of their detection efficiency.

If, on the other hand, one takes into account the trigger efficiency, as has been done by Aprile and Manzur, the systematic shift of the observed peak to higher energies should be removed.  However, a precise treatment of this procedure requires precise knowledge of the detection efficiency.  If the estimated detection efficiency is lower than the true efficiency, the reconstructed peak position will again become shifted.  Figure \ref{fig:Leff_Eff_Reconstruct_ex} (\emph{top}) shows the effect of compensating the observed spectrum with an incorrect estimate of the detection efficiency (purple curve); in particular, with one that is lower than the true efficiency.  In this case, the reconstructed peak position is \emph{lower} in energy than the true peak position.  Such an effect leads to a measured \Leff~value that is artificially lower than the true value, seen in Figure \ref{fig:Leff_Eff_Reconstruct_ex} (\emph{bottom}).  

The estimated detection efficiency used by Manzur is reported in \cite{Manzur:2009hp} as a function of the number of photoelectrons and that of Aprile is reported in \cite{ManalaysayThesis} as a function of $E_{ee}$.  \Leff~is reported as a function of $E_{nr}$.  In order to make a comparison, the two quantities (\Leff~and detection efficiency) can be independently scaled as a function of $E_{ee}$.  For \Leff, this is done by simply inverting equation \ref{eq:LeffDef}.  Figure \ref{fig:Leff_Eff_Reconstruct} shows this comparison for the results reported by Aprile and Manzur.  In the range of energies investigated by Aprile \emph{et al.}, the detection efficiency is near unity, dropping down to 97\% at the value with the lowest energy (5\,keV nuclear recoils).  For these data, therefore, systematic effects due to detection efficiency are minimal.  In contrast, the values reported by Manzur show a steep drop in \Leff~coinciding \emph{exactly} with their reported efficiency roll-off, so much so that the efficiency estimate almost appears to be a best fit when it is superimposed over the data, shown in Figure \ref{fig:Leff_Eff_Reconstruct}.  It seems unlikely that the laws of physics would produce a major change in \Leff~with exactly the same energy dependence as an important detector pathology inherent in the Manzur measurement.  This correlation indicates that the low energy behavior observed by Manzur is almost certainly an artifact of efficiency overcompensation.  A valid extrapolation of the low-energy behavior of \Leff~using the results of Manzur must therefore consider only those data with recoil energies greater than 10\,keV.

\section{Comparison of Indirect Measurements}
\label{sec:indirect_comparison}

The indirect method of determining \Leff, described in section \ref{sec:intro}, has been performed using data from the  XENON10 \cite{Sorensen:2008ec} and ZEPLIN-III \cite{Lebedenko:2008gb} detectors.  In these studies, \Leff~is parameterized as a function of nuclear recoil energy; the model parameters are then varied until the best fit is achieved between the observed and simulated neutron spectra.  As a cubic spline interpolation was used here in section \ref{sec:best_fit}, so was this method also used to parameterize \Leff~in \cite{Sorensen:2008ec}.  The model used in \cite{Lebedenko:2008gb}, however, is not stated, and appears to be a sigmoid function with likely two or three free parameters.  If this is the case, then the model is forced to fall sharply at decreasing energies, regardless of the parameter values.  Indeed, the high energy behavior of \Leff~reported in \cite{Lebedenko:2008gb} is perfectly flat, which is unsupported by all other measurements.

\begin{figure}[htp]
		\includegraphics[width=.48\textwidth]{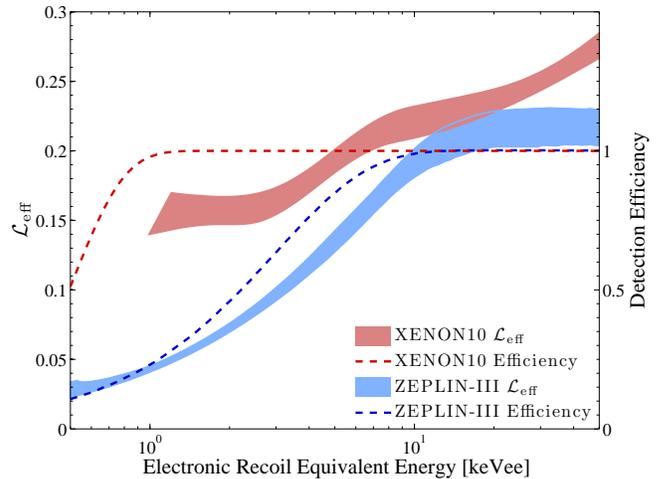}
	\caption{The reported \Leff~curves by XENON10 and ZEPLIN-III determined using the indirect method with a neutron source having a continuum energy spectrum (AmBe).  Superimposed are the energy dependent part of each detectors' efficiencies.  The energy independent efficiencies are 93.7\% and 59.1\% in XENON10 and ZEPLIN-III, respectively.}
	\label{fig:indirect_Leff_Effic}
\end{figure}

In section \ref{sec:beam_comparison}, comparison of the \Leff~energy dependence with the detection efficiency was extremely illustrative in diagnosing the observed results.  Here as well, the same comparison can be made.  However, while efficiency undercompensation and overcompensation had clearly predictable effects in direct beam measurements, the effect on indirect measurements is less obvious.  Undercompensation of trigger efficiency results in underpopulated histogram bins in the low energies, which will obviously disagree with the simulation.  However, the simulated spectrum can be made to fit this deficient measured spectrum, not by adjusting the bin heights according to the efficiency, but by suppressing \Leff~in this range.  Doing so pulls the low-energy simulated events below the energy threshold of the fit range.  In this way, depopulation of the low energy bins is done not by vertical scaling (efficiency compensation), but by horizontal scaling (energy compensation).  The result is a \Leff~curve that is artificially falling.

Figure \ref{fig:indirect_Leff_Effic}, analogous to Figure \ref{fig:Leff_Eff_Reconstruct}, shows the reported \Leff~curves from ZEPLIN-III \cite{Lebedenko:2008gb} (blue) and XENON10 \cite{Sorensen:2008ec} (red) as a function of electronic recoil equivalent energy with the reported detection efficiency curves superimposed.  Not only do the two \Leff~curves have virtually no overlap, they are made in regions where their efficiencies differ immensely.  The energy range of validity allowed by \cite{Sorensen:2008ec} sees an efficiency of no less than 99\%.  In contrast, the \Leff~curve reported in \cite{Lebedenko:2008gb} closely follows the measured efficiency roll-off.  This efficiency was not taken into account in \cite{Lebedenko:2008gb}, and therefore the reported \Leff~behavior is likely an artifact.  Indeed, if this represented the true behavior of \Leff, then the beam measurements discussed in section \ref{sec:beam_comparison} would see no signal at all in the low energies.

A possible problem of the results found by the indirect technique has recently been suspected \cite{Peter_davisTalk,Peter_priv}.  Neutrons from the source can scatter multiple times in LXe.  The time between interactions, $\mathcal{O}$(10\,ns), is similar to the time-scale of the scintillation emission, and therefore the multiple scintillation signals will be observed as a single pulse.  However, these multiple scatters are vetoed in time projection chambers (TPCs) like XENON10 and ZEPLIN-III by the observation of multiple ionization signals (separated by tens of $\mu$s).  In such detectors, however, there generally exist regions of LXe outside the active volume that are insensitive to charge, but nonetheless have a non-negligible optical coupling to the active LXe volume.  A neutron that scatters once in the active LXe volume, and once in a charge insensitive region will thus be reconstructed as a single scatter, with a scintillation signal that is anomalously too large since it is the sum of two interactions.

To some extent, these events can be removed from the data by cuts in energy and also the ratio of scintillation to ionization signals.  However, the effectiveness of these cuts is difficult to quantify, as is the optical coupling of the charge insensitive LXe regions.  Therefore, the spectra from data and simulation will not represent \emph{exactly} the same class of events.  But this pathology does not render the indirect measurements invalid.  The presence of multiple scatters affects mainly the high energy regions of the observed spectrum, because these are effects that only \emph{add} scintillation light.  A discrepancy in the high energies affects mainly the overall vertical offset of the reconstructed \Leff~curve.  However, the low energy shape of the result remains robust.

\section{Towards an Improved Understanding}
\label{sec:future}

\begin{figure}[htp]
		\includegraphics[width=.48\textwidth]{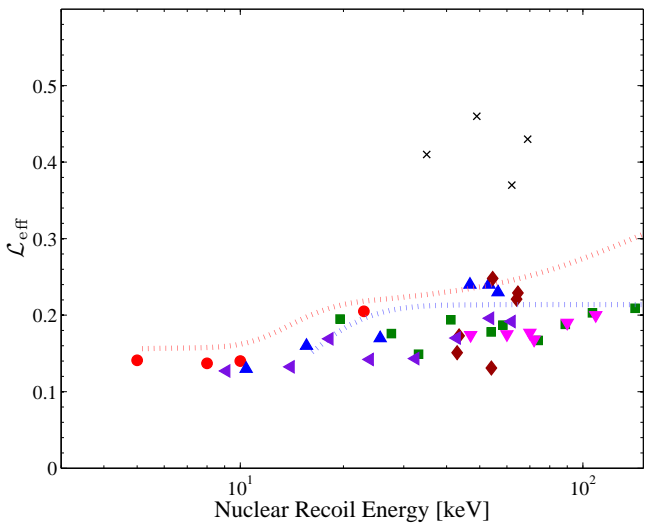}
	\caption{The landscape of \Leff~measurements, considering only results where the detection efficiency exceeds 90\% at the reported energy.  Additionally, the results of \cite{Manzur:2009hp} have been corrected using a more reasonable estimate of the neutron energy.  Markers and colors correspond to the same as those in Figures \ref{fig:Leff_fits} and \ref{fig:indirect_Leff_Effic}.}
	\label{fig:all_corrected}
\end{figure}

The rather disparate landscape of \Leff~measurements, both direct and indirect, casts doubt on our ability to currently resolve the underlying nature of this illusory quantity at low energies.  However, in sections \ref{sec:beam_comparison} and \ref{sec:indirect_comparison} it was shown that detection efficiency roll-off has, so far, \emph{always} produced sizable pathological effects on the determination of \Leff.  We can then use this understanding as a lens to focus our view of the landscape.  Figure \ref{fig:all_corrected} shows the available measurements with two modifications: (1) only measurements in regions where their detection efficiency is greater than 90\% are shown; (2) the results of \cite{Manzur:2009hp} have been corrected for a more reasonable estimate of the neutron energy.  Viewed in this way, a much clearer trend emerges: the combined results imply a \Leff~with no sharp changes with energy.  At 10\,keV, three measurements \cite{Aprile:05,Aprile:2008rc,Manzur:2009hp} show remarkable consistency of a value of $\mathcal{L}_{\rm eff}\simeq0.135$.

The behavior of \Leff~at energies below what are discussed here is difficult to predict.  Despite the appearance of equation \ref{eq:LeffDef}, \Leff~does not measure the ratio of scintillation response of nuclear to electronic recoils of the same energy.  This is because $L_y$, as it is used in the definition of \Leff, is valid only for electronic recoils at 122\,keVee, while it is known that the true electronic recoil $L_y$ grows at low energies \cite{Manalaysay:2009yq}.  If the true ratio of scintillation response from nuclear to electronic recoils drops at low energies (as it is expected to do), the nonlinearity of the electronic recoil light yield will act to soften that drop.  

The need for continued measurement of \Leff~is obvious, but what improvements can be made upon existing measurements?  The lessons of section \ref{sec:beam_comparison} are clear on this.  The spread in the true recoil energy must be minimized.  The contribution from the spread in the neutron energy is difficult to minimize, so instead one must focus on building a LXe chamber having a small volume, with the tagging scintillator(s) placed far away.  This must be coupled with a minimization of backgrounds due to multiple scatters in surrounding detector materials.  Light collection must be maximized, both to reduce effects from efficiency roll-off, but also to reduce uncertainties caused by energy resolution.

It appears that, although efficiency roll-off has so far caused problems, this systematic is one that cannot be completely avoided.  The measurements of Aprile \emph{et al.} \cite{Aprile:2008rc} were performed using a detector having virtually 4$\pi$ photo coverage of the LXe volume, which is why $L_y$ in that study (20\,pe/keVee) was so much larger than in any other similar study.  For this reason, the detection efficiency obtained in this study likely represents the practical limit in performance, and therefore any attempts at measurements below 5\,keV will necessarily involve effects due to trigger roll-off.  The emphasis in future studies, then, must be on a thorough and highly critical study of the efficiency roll-off, both with simulation and, more importantly, direct measurement.  A direct measurement can be done with the use of a high-energy gamma emitter such as $^{60}$Co (1.17\,MeV and 1.33\,MeV) or $^{208}$Tl (2.61\,MeV).  The Compton scatter spectrum of such sources is virtually flat, extending to zero energy \cite{Knoll:00}, so any significant deviations from flat behavior can be attributed to varying efficiency.  But regardless, the roll-off will likely carry a large systematic uncertainty, and we can therefore conclude that $\sim$5\,keV is the lowest energy at which \Leff~can be measured in the absence of an effect from efficiency roll-off.  Measurements of \Leff~below this energy can be accepted only after extremely careful scrutiny of the estimated detection efficiency; any significant disagreement between measurement and simulation of this efficiency should be treated with due skepticism.

\section{Summary}
\label{sec:summary}

The light yield of nuclear recoils in LXe, and its variation with energy, are neatly expressed as the parameter, \Leff.  It is rather unfortunate that, at energies most important for dark matter direct detection experiments, measurements of \Leff~largely disagree.  Unfortunate, but not unexpected, since such measurements are inundated with many systematic effects that can have large consequences for the results.  Sections \ref{sec:beam_comparison} and \ref{sec:indirect_comparison} exposed the most relevant of these systematic effects and showed not only that detection efficiency roll-off is the largest contributor to the disagreement between results, but exactly how this effect has produced the observed disparity.

When results are ignored that have been plagued by efficiency roll-off, the apparent disagreement disappears.  Seen in this way (Figure \ref{fig:indirect_Leff_Effic}), the available data are all consistent with a featureless \Leff, falling slowly from $\sim$0.2 at 100\,keV to $\sim$0.12 at 5\,keV.  Coincidentally, this behavior is very well approximated by the best fit from section \ref{sec:best_fit}.

\begin{acknowledgments}
I am grateful to Dr.~T.~Marrod\'an Undagoitia, Dr.~M.~Schumann, and Dr.~P.~Sorensen for reading this paper and providing comments, and to Prof.~L.~Baudis, Dr.~A.~Manzur, Prof.~D.~McKinsey, and Prof.~K.~Ni for many useful discussions on the topics contained within.  I also thank the XENON100 collaboration for using an early version of the best fit in their recent analysis \cite{Aprile:2010um}.
\end{acknowledgments}

\end{document}